\documentstyle[12pt]{article}
\begin{document}
\def\be{\begin{equation}}
\def\ee{\end{equation}}
\def\ba{\begin{array}}
\def\ea{\end{array}}
\parskip=6pt
\baselineskip=22pt
{\raggedleft{\sf{ ASITP}-94-9\\}}
{\raggedleft{\sf{~~ {February}} 1994.\\}}
\bigskip
\medskip
\centerline{\Large\bf  Matrix Realization of Gauge Theory
 on Discrete Group $Z_2$}
\vspace{10mm}
\centerline{\large \sf Jian-Ming Li}
\vspace{1.5ex}
\centerline{\sf CCAST (World Laboratory), P. O. Box 8730, Beijing 100080,
China;}
\vspace{1ex}
\centerline{\sf  Institute of Theoretical Physics, Academia Sinica,
P. O. Box 2735, Beijing 100080, China.\footnote{\sf Mailing address}\\}
\vspace{4ex}

 \centerline{Abstract}
{\it We construct a $2\times 2$ matrix algebra as representation  of functions
on discrete group $Z_2$ and develop the gauge theory on discrete group proposed
by Starz in  the matrix algebra. Accordingly, we  show that  the
non-commutative geometry model  built by R.Conquereax,
G.Esposito-Farese
and G.Vaillant results from this approach directly. For the purpose  of
Physical model building,
we introduce a free fermion Lagrangian on $M_4\times Z_2$ and study Yang-Mills
like gauge theory.}

\section[toc_entry]{Introduction}
Since Alian Connes introduced non-commutative geometry in particle physics to
explain the nature of Higgs field and the symmetry breaking mechanism[1], many
efforts have been done alone similar direction[2-6].  It is worthy to mention
two interesting approaches,
 one was  given by   R.Conquereax, G.Esposito-Farese
and G.Vaillant [3],  the other by Sitarz and the authors[6,10,11]. Both of them
are easy to be understood without entire knowledge of Non-commutative geometry.

Recently,CEV model has been investigated by some authors[7-9],
the main idea is  generalize the ordinary Yang-Mills potential by a finite
matrices. By defining an exterior algebra of forms over the non-commutative
direct product algebra of smooth function on spacetime and the Hermitian of the
$2\times 2$ matrices, a  Lagrangian can be constructed from consideration of
the generalized two forms. Another was proposed by Sitarz[6], who develop a
systematic approach towards the construction of pure gauge theory on arbitrary
discrete groups and  the authors[10,11] finished the physical model
building.

In this paper, based on the gauge theory on discrete group $Z_2$, we develop a
kind of  differential calculation on $2\times 2$ matrix algebra and show that
all the main formulas in CEV model may be derived from this approach. In other
words,
 the CEV model may be regarded as a matrix  representation of the later one.
Finally,  we discuss how to introduce the coupling between Higgs field and
fermion field.

\section[toc_entry]{Differential Calculation on $2\times 2$ Matrix Algebra}
Let $Z_2$ be a discrete  group  composed of  two elements  $Z_2=\{e, Z|
Z^2=e\}$
and ${\cal A}$ be the algebra of complex valued functions on $Z_2$.
 Every function f on discrete group $Z_2$ may be represented
as a $2\times 2$ diagonal matrix as follow
\be
{\cal F}=\left(\ba{ccc}f(e)&\\
&&f(Z) \ea\right).
\ee
We notice that all these matrices  construct a  algebra ${\cal M}$.
Using the result of differential calculus on discrete group $Z_2$, we
define $\bar{\partial}_Z$ on the matrices algebra as follow
\be
\bar{\partial}_Z=\left(\ba{ccc}\partial_Z&&\\ &&\partial_Z\ea\right)
\ee
where the action $\partial_Z$ on $f\in {\cal A}$ was defined by
$\partial_Zf=f-R_Zf$,  the right action of the group $R_Z$ acting on f as
$(R_Zf)(g)=f(g\odot Z)$, $g\in Z_2$ and $\odot$ is the group multiplication.

It is obviously that one dimension  space ${\cal E} $
which is spanned by $\bar{\partial}_Z$
forms an algebra
\be
\bar{\partial}_Z\cdot \bar{\partial}_Z=2\bar{\partial}_Z
\ee

Having the basis of ${\cal E}$, we can introduce the basis for the dual
space ${\cal E}^*$ or $\bar{\Omega}^1$
consisting of forms $\bar K$, which satisfy
\be
\bar K(\bar{\partial}_Z)=I.
\ee
The definition for the higher order forms is nature  and we take
$\bar{\Omega}^n $
to be the tensor product of n copies of $\bar{\Omega}^1$,
$\bar{\Omega}^n=\bar{\Omega}^{\otimes n}  $ and $\bar{\Omega}^0={\cal M} $.
To complete the construction of the differential algebra
$\bar{\Omega}^*=\{\displaystyle\bigoplus_n\bar{\Omega}^n\}  $, we need to
define the exterior
derivative $\bar d$, $\bar d: \bar{\Omega}^n\rightarrow \bar{\Omega}^{n+1}  $
whose
action on  ${\cal M}$ is defined by
\be
\bar{d} {\cal F}=\bar{\partial}_Z {\cal F} \bar K. \label {eq1}
\ee

If we realized $\bar d$ by a matrices operator
\be
\bar {d}=i\left(\ba{cc}0&d\\d&0\ea\right)
\ee
where operator d on algebra ${\cal A}$ as follow
\be
df=\partial_Z f \epsilon,
\ee
and $\epsilon$ is a unit vector, which construct a tensor algebra
$\Omega^*=\{\displaystyle\bigoplus_n \Omega^{n}\}, \Omega^0={\cal A},
\Omega^n=\Omega^{\otimes n}$.
Acting the operator ${\bar d}$ on ${\cal F}\in {\cal M}$, we have
\be
\bar{d}{\cal F}=i\left(\ba{ccc}&&df(e)\\
df(Z)&&\ea\right)=i\bar{\partial}_Z {\cal
F}\left(\ba{ccc}&&\epsilon\\\epsilon&&\ea
\right).\label {eq2}\ee
Comparing eq.(\ref {eq1}) with (\ref {eq2}),
 we  get the matrices representation
of $\bar K$ with $\epsilon$
  $$\bar K= \left(\ba{ccc}&&i\epsilon\\i\epsilon&&\ea\right).$$

If we define matrix tensor product as follow
\be
\left(\ba{ccc}A&B\\C&D\ea\right)\otimes
\left(\ba{ccc}A'&B'\\C'&D'\ea\right)=
\left(\ba{ccc}A\otimes A'+B\otimes C'&
A\otimes B'+B\otimes D'\\C\otimes A'+D\otimes C'&C\otimes B'+D\otimes
D'\ea\right),\ee
where $A\cdots D' \in \Omega^*$.
We get the matrices representation of forms ${\bar K}^n={\bar K}^{\otimes n}$
with
${\epsilon}^{\otimes n}$ as follows:
\be
{\bar K}^{2n}=\left(\ba{ccc}(i\epsilon) ^{2n}&&\\&&(i\epsilon)^{2n}\ea\right),
{\bar
K}^{2n-1}=\left(\ba{ccc}&&(i\epsilon)^{2n-1}\\(i\epsilon)^{2n-1}&&\ea\right).
\ee

To complete the construction of the differential algebra we need to define
the exterior derivative $\bar d$ and this is the subject of the following
lemma:

{\it There exists exactly one linear exterior derivative operator $\bar d$
such that it satisfies
\be\ba{cl}
(i)&~~\bar{d}^2=0,\\
(ii)&~~\bar{d}(A\otimes B)=\bar{d}A\otimes B+(-1)^{Deg A}A\otimes
\bar{d}B,~~\forall A,B \in \bar{\Omega}^*\\
(iii)&~~(\bar{d}{\cal F})(V)=V({\cal F}),~~ \forall V \in {\cal E}, {\cal F}\in
{\cal M}
\ea\ee
provided that $\epsilon$ satisfies the following conditions
\be\ba{cl}
&\epsilon f=f\epsilon, ~~\forall f \in {\cal A}\\
&d\epsilon=-2\epsilon \otimes \epsilon.
\ea\ee}

Using the Lemma, it is easy to show that
$$d\bar K^{2n}=0,~~~d\bar K^{2n-1}=-2 \bar K^{2n}.$$
Then we have
\be\ba{cl}
\bar{d} ({\cal F}\bar{K}^{2n})&=\bar d{\cal F}\bar K^{2n}+{\cal F}\bar d \bar
K^{2n}\\[4mm]
&=\left(\ba{ccc}&(f(e)-f(Z))(i\epsilon)^{2n+1}\\(f(Z)-f(e))
(i\epsilon)^{2n+1}&\ea\right)\\[10mm]
\bar{d} ({\cal F}\bar{K}^{2n-1})&=\bar{d}{\cal F}\bar K^{2n-1}+{\cal F}\bar d
\bar
K^{2n-1}\\[4mm]&=\left(\ba{ccc}-(f(e)+f(Z))(i\epsilon)^{2n}&\\&-(f(Z)+f(e))
(i\epsilon)^{2n}\ea\right),\ea\ee
which corresponding to the derivative on even and odd matrices of CEV model
\be
da_e=i \left(\ba{ccc}&&a_{22}-a_{11}\\a_{11}-a_{22}&&\ea\right),~~
da_o=i \left(\ba{ccc}a_{21}+a_{12}&&\\&&a_{21}+a_{12}\ea\right).\ee

The involution on the differential algebra agrees with the complex
conjugation ${\cal M}$ and commutes with $\bar d$, i.e. $\bar
{d}(\bar\omega)^*=(-1)^{Deg\bar {\omega}} (\bar d \bar \omega)^*$. Again it
is sufficient to calculate it if we set the involution $\bar K$, the basis
of the one forms, we have $\epsilon^*=\epsilon$.

Let us now construct the generalized gauge theory using the above
differential forms. We take the gauge transformations to be any proper
subset ${\cal H} \subset {\cal M}$. In particular, we will often take
${\cal H}$ to be the group of unitary elements of ${\cal M}$
\be {\cal H}={\cal U}({\cal M})=\{a\in {\cal M}: aa^{\dag} =a^{\dag} a=I\}.
\ee
It is easy to see that the exterior derivative $\bar d$ is not covariant with
respect to the gauge transformations so that we should introduce the
covariant derivative $\bar d+\bar {\phi}$, where $\bar \phi=\phi \bar K,
\phi\in {\cal M}$ is generalized
connection one-form. The requirement that $\bar d +\bar \phi $
is gauge invariant under the gauge transformations
\be
\bar d +\bar\phi=H(\bar d +\bar\phi)H^{-1}, H\in {\cal M},\ee
results in the following transformation rule of $\bar \phi$
\be
\bar \phi\rightarrow H\bar \phi H^{-1} +H\bar d H^{-1}\ee
and $\phi$ transform as
\be
\phi \rightarrow H\phi\left(\ba{ccc}&&1\\1&&\ea\right) H^{-1}
\left(\ba{ccc}&&1\\1&&\ea\right)+H\bar{{\partial}_Z }H^{-1}\label {a}.\ee
It is convenient  to introduce a new field $\Phi=1-\phi$, then (\ref a)
is equivalent to
\be
\Phi\rightarrow H\Phi \left(\ba{ccc}&&1\\1&&\ea\right) H^{-1}
\left(\ba{ccc}&&1\\1&&\ea\right).
\ee
It can be shown that the generalized curvature two form
\be
\bar{F}=\bar {d}\bar \phi +\bar\phi\otimes \bar{\phi}
\ee
is gauge invariant.

In the following, we set $\bar\phi=\left(\ba{ccc}\phi(e)&&\\&&\phi(Z)\ea\right)
\cdot\bar  K=\left(\ba{ccc}&&i\phi(e)\epsilon\\i\phi(Z)\epsilon&&\ea
\right)   $. The condition ${\bar \phi}=-{\bar \phi}^{\dag}$ enforce the
following relation of its coefficients.
\be
\phi(Z)={\phi(E)}^{\dag}.\ee

Then we have
\be
\bar F=\left(\ba{ccc}-(\Phi\Phi^{\dag}-1)\epsilon\otimes
\epsilon&&\\&&-(\Phi^{\dag}\Phi-1)\epsilon
\otimes \epsilon\ea\right).
\ee

In order to construct the Lagrangian of the gauge theory we need to
introduce a metric. Let us define
\be
<\epsilon,\epsilon> =\eta,~~<\epsilon\otimes \epsilon,\epsilon\otimes
\epsilon>= <\epsilon,\epsilon> <\epsilon,\epsilon> =\eta^{2},\ee
then we have the Yang-Mills like Lagrangian of the gauge field
\be
{\cal L}=-||F||^2 =-Tr<F,F>=-|F_{11}|^2-|F_{12}|^2-|F_{21}|^2- |F_{22}|^2
=-2\eta^2(\Phi\Phi^{\dag}-1)^2
\ee

\section[toc_entry]{Gauge Theory on  $ M^4\times Z_2  $ }

In order to get a full  Lagrangian of the Higgs field, the kinetic term must
be include. To this end, we extend the exterior derivative operator ${\bar
d}_{Z_2}$ to the one as follows:
\be
\bar{d} = \bar{d}_M+\bar{d}_{Z_2}\ee
where $\bar{d}_M=\left(\ba{ccc}d_M&&\\&&d_M\ea\right)$, $d_M$ is the
exterior derivative on spacetime $M^4$. The nilpotency of $\bar{d}$
requires that
\be
\bar{d}_M\bar{d}_{Z_2}=-\bar{d}_{Z_2}\bar{d}_M.\label b\ee
Acting $\bar{d}_M\bar{d}_{Z_2}$ and $\bar{d}_{Z_2}\bar{d}_M$ on $f(x)\in {\cal
M}$ separately, relation (\ref b) is equivalent to
\be
\partial_\mu\partial_{Z_2}f(x,g)dx^\mu\otimes
\epsilon=-\partial_{Z_2}\partial_\mu
f(x,g) \epsilon\otimes dx^\mu,~~~g\in Z_2
\ee
thus we have
\be
dx^\mu\otimes \epsilon=-\epsilon\otimes dx^\mu.
\ee

The most general connection one-form on $M^4\times Z_2$ can be written as
$$\bar A={\bar A}_\mu dx^\mu+\bar \phi \bar K=
\left(\ba{ccc}A_\mu(x,e)dx^{\mu}&i\phi\epsilon\\i\phi^{\dag}\epsilon&A_\mu(x,Z)dx^\mu\ea
\right),$$ Using the exterior derivative operator defined by (24), we have
\be\ba{cl}
\bar{d}\bar A
&=\left(\ba{ccc}\partial_\mu A_\nu dx^\mu\wedge dx^\nu
+(\phi+\phi^{\dag})\epsilon\otimes
\epsilon&i(\partial_\mu\phi-(A_\mu-B_\mu))dx^\mu\otimes \epsilon\\
i(\partial_\mu\phi^{\dag}+(A_\mu-B_\mu ))dx^\mu\otimes \epsilon&
\partial_\mu A_\nu dx^\mu\wedge dx^\nu+(\phi+\phi^{\dag})\epsilon\otimes
\epsilon \ea
\right) \\[6mm]
\bar A\otimes \bar A&=\left(\ba{ccc} A_\mu dx^\mu\wedge A_\nu dx^\nu
-\phi\phi^{\dag}\epsilon\otimes
\epsilon&i(A_\mu\phi-\phi B_\mu)dx^\mu\otimes \epsilon\\
i(B_\mu\phi^{\dag}-\phi^{\dag}A_\mu)dx^\mu\otimes \epsilon&
 B_\mu dx^\mu\wedge B_\nu dx^\nu-\phi\phi^{\dag}\epsilon\otimes \epsilon\ea
\right)\ea.\label {n}\ee
where $A=A(x,e), B=A(x,z)$.

The formulas (\ref n) are equavalent to formulas in CEV model,
if we write ${\cal A}=\left(\ba{ccc}A&-i\phi\\-i\phi^{\dag}&B\ea\right)$
$$\bar d{\cal A}=\left(\ba{ccc}dA
+(\phi+\phi^{\dag})
&i(d\phi-(A-B))\\
i(d\phi^{\dag}+(A-B))&
dB+(\phi+\phi^{\dag})\ea\right), \label {m}$$

$${\cal A}\otimes {\cal A}=\left(\ba{ccc} A\wedge
A-\phi\phi^{\dag}&i(A\phi-\phi B)\\
i(B\phi^{\dag}-\phi^{\dag}A)&
 B\wedge B-\phi\phi^{\dag}\ea\right).$$

The generalized curvature two form is
$$F=\bar{d}\bar{A}+\bar{A}\otimes
\bar{A}=\left(\ba{ccc}F_{11}&F_{12}\\F_{21}&F_{22}\ea\right),$$
from the above calculation we have
$$\ba{cl}
&F_{11}=F_{\mu\nu}(x,e)dx^\mu\wedge dx^\nu-(\Phi\Phi^{\dag}-1)\epsilon
\otimes \epsilon\\
&F_{12}=-iD_\mu\Phi dx^\mu\otimes \epsilon \\
&F_{21}=-iD_\mu\Phi^{\dag}dx^\mu\otimes \epsilon\\
&F_{22}=F_{\mu\nu}(x,Z)dx^\mu\wedge dx^\nu-(\Phi\Phi^{\dag}-1)\epsilon
\otimes \epsilon,\ea$$
where
$$F_{\mu\nu}=\partial_\mu A_\nu-\partial_\nu A_\mu+[A_\mu,A_\nu],$$
$$D_\mu\Phi=\partial \Phi+A_\mu (x,e)\Phi-\Phi A_\mu(x,Z).$$
$$\Phi=1-\phi$$

To get the lagrangian of the pure gauge field, we define the metric as
\be\ba{cl}
<dx^\mu,dx^\nu>=g^{\mu\nu}&,~~<\epsilon,\epsilon>=\eta\\
<dx^\mu\otimes \epsilon, dx^\nu\otimes
\epsilon>=g^{\mu\nu}\eta&,~~<\epsilon\otimes
\epsilon,\epsilon\otimes \epsilon>=\eta^2\ea\ee

Using the previous metric defination, we can get the largrangian for the
gauge field:
\be
{\cal L}=\frac 1 N \{-\frac 1 4 TrF_{\mu\nu}F^{\mu\nu}(x,e)-\frac 1 4
TrF_{\mu\nu}F^{\mu\nu}(x,Z)+\eta TrD_\mu\Phi D^\mu\Phi^{\dag}-\eta^2
Tr(\Phi\Phi^{\dag}-1)^2\}
\label {x}.\ee

Sometime it's useful to calculate with Dirac gamma matrices.
We define a vector space isomorphism map $\pi$
from the exterior algebra and the Clifford algebra,
$$\pi:~dx^\mu\rightarrow \gamma^\mu, \epsilon\rightarrow \gamma^5$$
This is necessary when we want to couple the gauge fields to
the spinors.

{}From previous discussion, the results from this aproach is very similar  to
those in CEV model, except a form basis $\epsilon$ is introduced here.

\section[toc_entry]{Fermion}
A Hilbert space $\bar H$ is composed of  Dirac spinnor, a vector  $
\Psi\in \bar H$ is defined as follow
\be
\Psi=\left(\ba{c}\psi(x,e)\\\psi(x,Z)\ea\right)
=\left(\ba{c}\psi_L(x)\\\psi_R(x)\ea\right).\ee
where
$$\psi(x,e)={\psi}_{L}(x)=\frac{1}{2}(1+{\gamma}_{5}) \psi(x),
{}~~~\psi(x,Z)={\psi}_{R}(x)=
\frac{1}{2}(1-{\gamma}_{5}) \psi(x)$$
are the left and right handed Dirac field respectively.

Then the lagrangian for the free fermion can be written as
\be
{\cal L}_F=\overline{\Psi}i\bar{\gamma}^i\bar{\partial}_i\Psi\label {e}, \ee
where
\be\ba{cl}
&\bar{\gamma}^i=\left\{\ba{cl}&\left(\ba{ccc}\gamma^\mu&\\&\gamma^\mu\ea\right),
\mu=0,1,2
,3,\\[4mm]
&\left(\ba{ccc}i\gamma^5& \\&-i\gamma^5\ea\right),~~i=5;\ea\right.\\[12mm]
&\bar{\partial}_i=\left\{\ba{cl}&\bar{\partial}_\mu,~~i=\mu=0,1,2,3,\\[4mm]
&\mu\bar{\partial}_Z,~~i=5,\ea \right.\ea\ee
the reason  is  that
$${\cal L}_F=\overline{\psi}(x)(i{\gamma}^\mu-\mu)\psi(x).$$
If we require the lagrangian (\ref e) is invariant under the gauge
transformation
$U\in {\cal M}$, we should introduce the covariant derivative $D_\mu, D_Z$,
where $\bar D_\mu=\bar\partial_\mu+igA_\mu$,$\bar D_Z=\bar\partial_Z+\frac
\lambda \mu \phi\bar R_Z$,
the convariant derivative on discrete group is definite by the convariant
differential forms,
$$\bar{D}_Zf=(\bar{\partial}_Z+\frac \lambda \mu\phi
\bar{K})f=(\bar{\partial}_Z+
\frac \lambda \mu\phi\bar{R}_Z)f\bar{K},~~f,\phi\in {\cal M},$$ where $\bar
R_Z=\left(\ba{ccc}R_Z&\\&R_Z\ea\right)$.
 Then we have $$\bar{D}_Z=\bar{\partial}_Z+\frac\lambda \mu \phi\bar{R}_Z$$ and
the lagrangian (\ref e) can be written as
\be\ba{cl}
{\cal L}_F=&\overline{\psi}_L(x)i\gamma^\mu
D_{L\mu}\psi_L(x)+\overline{\psi}_R(x)i\gamma^\mu D_{R\mu}\psi_R(x)\\
&-\lambda\overline{\psi}_L(x)\Phi(x)\psi_R(x)-
\lambda\overline{\psi}_R(x)\Phi^{\dag}(x)\psi_L(x),\ea\ee
where $\Phi=\frac \mu \lambda-\phi$, and all the coupling in the lagrangian are
gauge coupling.

Noticing the fact that
$$\bar{\gamma}^i {\bar D}_i=\bar
{\gamma}^i\bar{\partial}_i+\left(\ba{ccc}ig_1\gamma^\mu
A_\mu(x,e)&i\lambda\gamma^5\phi\\-i\lambda\gamma^5\phi^{\dag}&ig_2\gamma^\mu
A_(x,Z)\mu\ea\right)
$$  and using the vector space isomorphism of the exterior algebra and Clifford
algebra, we have
$$\pi^{-1}:~~\left(\ba{ccc}ig_1\gamma^\mu
A_\mu(x,e)&i\lambda \gamma^5\phi\\-i\lambda \gamma^5\phi^{\dag}&ig_2\gamma^\mu
A_(x,Z)\mu\ea\right)
\rightarrow\left(\ba{ccc}ig_1
A_\mu(x,e)dx^\mu&\lambda
i\phi\epsilon\\\lambda(i\phi)^{\dag}\epsilon&ig_2A_\mu(x,Z)dx^\mu\ea\right).$$
Therefore, we introduce the gauge invariant lagrangian for the Boson field
parts ${\cal L}_{Boson}$ from(\ref x)
\be
{\cal L}_{Boson}=\frac 1 N \{-\frac {g_1^2} 4 TrF_{\mu\nu}F^{\mu\nu}(x,e)-\frac
{g_2^2} 4
TrF_{\mu\nu}F^{\mu\nu}(x,Z)+\eta \lambda^2 TrD_\mu\Phi D^\mu\Phi^{\dag}-\eta^2
\lambda^4 Tr(\Phi\Phi^{\dag}-\frac {\mu^2 }{\lambda^2})^2\}
\ee
and the complete lagrangian is then ${\cal L}={\cal L}_{Fermion}+{\cal
L}_{Boson}$.

To conclude, we emphasize that both the two approaches are originate  from  the
work of Connes[1]. In the following paper we will show that gauge theory on
discrete group  $Z_2$ equavalent to Connes approach for two discrete points.

\bigskip

\centerline{\large \bf {Acknowledgements}}

\noindent  The author would like to thank  Professors H.Y Guo, K. Wu and Dr. Y.
K. Lau
for helpful discussions.

\newpage

\centerline{\large  {References}}
\bigskip

\begin{enumerate}

\item A. Connes, in: The Interface of  Mathematics and Particle Physics,\\
eds. D. Quillen, \- G. Segal and S. Tsou \- (Oxford U. P, Oxford 1990);
\-A. Connes and J. Lott, \- Nucl. Phys. (Proc. Suppl.) {\bf B18}, 44 (1990);
\-A. Connes and Lott, \- Proceedings of 1991 \-Cargese Summer \- Conference;
\- See also A. Connes, \-{\it Non-Commutative \- Geometry.}~~{\bf
IHES/M}/93/12.

\item D. Kastler, Marseille, CPT preprint {\bf CPT-91}/P.2610, {\bf
CPT-91}/P.2611.

\item R. Coquereaux, G. Esposito-Far\'ese and G Vaillant, Nucl Phys {\bf B353}
 689 (1991).

\item M. Dubois-Violette, \-  R. Kerner, J. Madore, J. Math. Phys.
{\bf 31}. (1990) 316.

\item A. H. Chamseddine, \- G Felder and J. Fr\"ohlich, \- Phys. Lett. \-
 {\bf 296B} (1993) 109.

\item A. Sitarz,  Non-commutative Geometry and  Gauge Theo\-ry on Dis\-crete
Groups, preprint {\bf TPJU}-7/1992;   A.Sitarz, Phys. Lett {\bf 308B}(1993)
311.

\item B. S. Balakrishna, F G\"ursey and K. C. Wali, \-  Phys Lett {\bf B254},
430 (1991).

\item R. Coquereaux, G. Esposito-Far\'ese, Int. J. Mod. Phys, {\bf A7}
(1992)6555.

\item B.E.Hanlon and G.C.Joshi, Lett. Math.Phys,{\bf 27}(1993)105.

\item Haogang Ding, Hanying Guo, Jianming Li and Ke Wu,
 Commun. Theor. Phys. {\bf
21} (1994) 85-94; J. Phys. A:Math.Gen.  {\bf 27}(1994) L75-L79;  J. Phys.
A:Math.Gen.  {\bf 27} (1994)L231-L236.

\item Hao-Gang Ding, Han-Ying Guo, Jian-Ming Li and Ke Wu, \- Higgs as Gauge
Fields On Discrete \-Groups and Standard Model For \- Electroweak and
Electroweak-Strong Interactions, \-preprint {\bf ASITP}-93-23,{\bf CCAST}-93-5,
To appear in Z. Physik C.

\end{enumerate}
\end{document}